\documentclass[trackchanges]{aastex7}

\usepackage{float}
\usepackage{graphicx}
\usepackage{natbib}
\usepackage{multirow} 
\usepackage{epstopdf}
\shorttitle{Herringbone structures during an X-class eruptive flare}
\shortauthors{Zhang et al.}

\begin{document}
	
	\title{Herringbone structures during an X-class eruptive flare}
	
	\correspondingauthor{Qingmin Zhang}
	\email{zhangqm@pmo.ac.cn}
	
	\author[0000-0003-4078-2265]{Qingmin Zhang}
	\affiliation{Purple Mountain Observatory, Chinese Academy of Sciences, Nanjing 210023, People's Republic of China}
	\email{zhangqm@pmo.ac.cn}
	
	\author[0000-0002-9893-4711]{Zongjun Ning}
	\affiliation{Purple Mountain Observatory, Chinese Academy of Sciences, Nanjing 210023, People's Republic of China}
	\email{ningzongjun@pmo.ac.cn}
	
	\author[0000-0002-1810-6706]{Xingyao Chen}
	\affiliation{State Key Laboratory of Solar Activity and Space Weather, National Space Science Center, 
	Chinese Academy of Sciences, Beijing 100190, People's Republic of China}
	\email{xingyao.chen@njit.edu}
	
	\author[]{Wei Chen}
	\affiliation{Purple Mountain Observatory, Chinese Academy of Sciences, Nanjing 210023, People's Republic of China}
	\email{w.chen@pmo.ac.cn}
	
	\author[0000-0003-2891-6267]{Xiaoli Yan}
	\affiliation{Yunnan Observatories, Chinese Academy of Sciences, Kunming 650216, People's Republic of China}
	\affiliation{Yunnan Key Laboratory of Solar Physics and Space Science, Kunming 650216, People's Republic of China}
        \email{yanxl@ynao.ac.cn}
	
	\author[0009-0000-5839-1233]{Shuyue Li}
	\affiliation{Purple Mountain Observatory, Chinese Academy of Sciences, Nanjing 210023, People's Republic of China}
	\affiliation{School of Science, Nanjing University of Posts and Telecommunications, Nanjing 210023, People's Republic of China}
	\email{syli@pmo.ac.cn}

\begin{abstract}
In this paper, we report quasi-periodic herringbone structures during the impulsive phase of an X-class flare, 
coinciding with the distinct acceleration phase of eruptive prominence ejection on 2023 December 31. 
The prominence propagates non-radially in the southeast direction with an inclination angle of $\sim$35$\fdg$4.
The fast coronal mass ejection (CME) at a speed of $\sim$2852 km s$^{-1}$ drives a shock wave and a coronal EUV wave.
The herringbone structures lasting for $\sim$4 minutes take place at the initial stage of a group of type II radio burst.
The herringbones in the frequency range 20$-$70 MHz are characterized by simultaneous forward-drift and reverse-drift bursts
with average durations of $\sim$2.5 s and $\sim$3.1 s.
The frequency drift rates of these bursts fall in a range of 1.3$-$9.4 MHz s$^{-1}$ with average values of $\sim$3.6 and $\sim$4.1 MHz s$^{-1}$, respectively. 
The speeds of electron beams producing the herringbones are estimated to be 0.04$-$0.41 $c$, 
with average values of $\sim$0.23 $c$ and $\sim$0.11 $c$ for forward-drifting and reverse-drift bursts, respectively.
The heights of particle acceleration regions are estimated to be 0.64$-$0.78 $R_{\sun}$ above the photosphere,
which are consistent with the height of CME front ($\sim$0.75 $R_{\sun}$) when the shock forms.
Quasi-periodic pulsations with periods of 17.5$-$21.3 s are found in the radio fluxes of herringbones,
suggesting that electrons are accelerated by the CME-driven shock intermittently.
\end{abstract}
	

\section{Introduction} \label{intro}
Solar flares and coronal mass ejections (CMEs) are the most spectacular eruptions as a result of impulsive energy releases in the solar system \citep{fle11}.
Quasi-periodic pulsations (QPPs) could be observed in soft X-ray (SXR), hard X-ray (HXR), extreme ultraviolet (EUV), UV, 
and radio wavelengths \citep{ing08,cxy19,zim21,lsy25}.
QPPs may reflect intermittent magnetic reconnections and particle accelerations in flares \citep{zqm16}.
Fast CMEs are likely to drive shock waves, which are usually associated with type II radio bursts in the metric and decameter wave ranges \citep{mann95a,vrs02,zuc14}.
The frequencies of a type II burst drift slowly from high to low values as the CME propagates outward.
Like termination shocks generated by reconnection outflow jets in solar flares \citep{aur02,chen15}, 
CME-driven shocks are capable of accelerating particles to high energies as well \citep[see][and references therein]{zank07}.
\citet{gop09b} investigated CMEs during the declining phase of solar cycle 23 and derived the CME heights when the related type II bursts start.
The average distance is $\sim$1.5 $R_{\sun}$, exactly consistent with the location at which the Alfv\'{e}n speed reaches a minimum.
Besides, the CME-driven shocks are most likely to accelerate electrons in the heliocentric distance of 1.5$-$4.0 $R_{\sun}$.
\citet{gop13} obtained the shock formation height ($\sim$1.38 $R_{\sun}$) and the starting frequency ($\sim$85 MHz) 
of a type II burst related with the CME on 2012 May 17.
\citet{zuc14} reported that the shock forms at the CME flank with a height of $\sim$1.6 $R_{\sun}$ on 2011 September 23.

Herringbones are fine structures related to type II radio bursts \citep{rob59,cai87,cane89,mann95a,mann05,doro15,kov23,mor24}.
They are composed of simultaneous forward-drift and reverse-drift radio bursts in most cases \citep{mann92}.
Sometimes, herringbones show merely forward-drift or reverse-drift bursts \citep{cai87,mann95b,mag20}.
The drift rates of herringbones are a few MHz s$^{-1}$, which are much faster than typical type II bursts ($\sim$0.1 MHz s$^{-1}$).
In a statistical study, \citet{mann02} found that the drift rates of herringbones are nearly half of those of type III bursts.
Surprisingly, \citet{mag20} reported a drift rate up to $\sim$25 MHz s$^{-1}$ of the herringbones on 2014 August 25.
The durations of herringbones are mostly a few minutes \citep{mann92,car13,car15,mor19,car21} and rarely up to $\sim$20 minutes \citep{mor22}.
It is generally believed that herringbones are produced by high-energy electron beams repetitively accelerated by quasi-perpendicular CME-driven shocks,
with the shock normal angle ($\theta_{\mathrm{BN}}$) between the shock normal and the upstream magnetic field being close to 90$\degr$ \citep{bale99}.
Specifically, these electrons are accelerated via shock drift acceleration at the wavy or rippled shock fronts 
so that electrons could be trapped and accelerated repeatedly to obtain very high energies \citep{hol83,zlo93,ball01,mann05,bur06,mit07,guo10,van11,mor24}.
The emission mechanism of herringbones is plasma emission \citep{dulk85}.
Consequently, propagations of these electrons toward and away from the Sun generate reverse-drift and forward-drift bursts, respectively.
It is demonstrated that forward-drift herringbone shows an opposite sense of circular polarization to the reverse-drift herringbone \citep{mor22}.
Analysis of herringbone structures provides an efficient and useful tool to probe plasma turbulence in the corona \citep{car21,kov23}.

Till now, herringbones have rarely been detected. It is found that only $\sim$20\% of type II bursts are associated with herringbones \citep{cai87,mag20,car21,mor22,mor24}.
On 2011 September 22, a hot channel eruption occurred in NOAA active region (AR) 11302, 
producing an X1.4 class flare and a fast CME at a speed of $\sim$1905 km s$^{-1}$ \citep{zuc14,zqm23}.
A coronal shock and an extreme ultraviolet (EUV) wave are driven by the CME, 
which is associated with herringbone structures \citep{car13} detected by spectrometers at the e-Callisto\footnote{https://www.e-callisto.org} \citep{benz09}
Rosse Solar Terrestrial Observatory \citep[RSTO;][]{zuc12}.
The herringbones, whose frequencies lie in the range of 30$-$80 MHz, last for $\sim$2 minutes.
The Alfv\'{e}n Mach number ($M_a$) of the shock reaches $\sim$2.4 and the kinetic energies of shock-accelerated electrons reach 2$-$46 keV (0.1$-$0.4 $c$),
where $c$ denotes the speed of light. Interestingly, the herringbone structures present a quasi-periodicity of 2$-$11 s.
\citet{mor19} studied the X8.2 class eruptive flare occurring in AR 12673 on 2017 September 10. 
The extremely fast CME at a speed of $>$3000 km s$^{-1}$ drives a shock wave ($M_{a}\sim2.9$), 
which accelerates electrons at multiple locations along the expanding CME.
Three groups of herringbones during five minutes are detected with the LOw-Frequency ARray \citep[LOFAR;][]{van13} in the frequency range 30$-$60 MHz.
The speeds of electron beams reach up to 0.20$-$0.25 $c$.

On 2023 December 31, an X5.0 class flare occurred in NOAA AR 13536 (N04E79) close to the eastern limb.
The eruptive flare was associated with a fast CME at a speed of $\sim$2852 km s$^{-1}$ recorded in the CDAW CME 
catalog\footnote{https://cdaw.gsfc.nasa.gov/CME\_list/} \citep{gop09a}.
Using the joint high-cadence and high-resolution observations from the Owens Valley Radio Observatory\rq{}s
Expanded Owens Valley Solar Array \citep[EOVSA;][]{gary18} and Long Wavelength Array \citep[OVRO-LWA;][]{and18},
\citet{cxy25} measured the magnetic field of the CME (magnetic flux rope) from low to middle corona. 
It is revealed that the magnetic field strength decreases rapidly from $\sim$300 G at 1.02 $R_{\sun}$ to $\sim$0.6 G at 2.83 $R_{\sun}$.
Meanwhile, magnetic flux conservation of the CME is confirmed.
In this paper, we focus on the quasi-periodic herringbone structures during the impulsive phase of the X-class flare, 
which are closely related to the distinct acceleration phase of a prominence eruption.
The paper is organized as follows. 
Section~\ref{obs} describes the instruments we use in this study.
Section~\ref{ep} shows the flare, CME, and EUV wave as a result of a prominence eruption.
Section~\ref{hb} shows the herringbone structures associated with the shock. 
Finally, a brief summary and discussions are given in Section~\ref{sum}.

\begin{deluxetable*}{ccccccc}
		\digitalasset
		\tablewidth{\textwidth}
		\tablecaption{Properties of the instruments used in this study.
			\label{tab1}}
		\tablecolumns{4}
		\tablenum{1}
		\tablehead{
			\colhead{Instrument} &
			\colhead{Waveband} &
			\colhead{Cadence} &
			\colhead{Pixel Size}  \\
			\colhead{ } &
			\colhead{ } &
			\colhead{(s)} &
			\colhead{(arcsec)} &
		}
		\startdata
		 SDO/AIA & 171, 211, 304 {\AA} & 12 & 0.6 \\
		 LASCO-C2 & WL & 720 & 11.4 \\
		 LASCO-C3 & WL & 720 &  56.0 \\
                  STA/COR2 & WL & 900 & 14.7  \\
                  GOES-16/SUVI & 195 & $\sim$60 & 2.5 \\
                  GOES-16/SUVI & 304 & $\sim$120 & 2.5 \\
		 GOES-18/XRS & 0.5$-$4 {\AA}, 1$-$8 {\AA} & 1 & -  \\
		 S/WAVES & 0.0025$-$16.025 MHz & 60 & - \\
		 ALASKA & 5$-$89 MHz & 0.25 & - \\
		 AUSTRALIA & 15$-$87, 108$-$377 MHz & 0.25 & - \\
		 MEXICO-LANCE-B & 45$-$90 MHz & 0.25 & - \\
		\enddata
\end{deluxetable*}

\section{Instruments and Observations} \label{obs}
The X5.0 class flare and associated CME were observed by several telescopes, 
including the Atmospheric Imaging Assembly \citep[AIA;][]{lem12} on board the Solar Dynamics Observatory \citep[SDO;][]{pes12},
the X-Ray Sensor \citep[XRS;][]{han96} on board the Geostationary Operational Environmental Satellite 18 \citep[GOES-18;][]{gar94} spacecraft, 
the Solar Ultraviolet Imager \citep[SUVI;][]{tad19,dar22} on board the GOES-16 spacecraft, 
the C2 and C3 coronagraphs of the Large Angle Spectroscopic Coronagraph \citep[LASCO;][]{bru95} on board the Solar and Heliospheric Observatory (SOHO) mission,
and the COR2 coronagraph of the Sun Earth Connection Coronal and Heliospheric Investigation \citep[SECCHI;][]{how08}
on board the Solar TErrestrial RElations Observatory \citep[STEREO;][]{kai08} ahead (hereafter STA).
The separation angle between STA and the Sun-Earth connection was 6$\fdg$9 \citep[see Fig. 2 in][]{ryan24}.
Type III and type II radio bursts associated with the eruption are identified in the radio dynamic spectra
observed by the S/WAVES \citep{bou08} on board STA and e-Callisto AUSTRALIA.
The herringbone structures were recorded by the e-Callisto AUSTRALIA, ALASKA, and MEXICO-LANCE-B.
The properties of instruments and their data are summarized in Table~\ref{tab1}.

\section{Results} \label{res}
\subsection{Flare, CME, and EUV wave} \label{ep}
\begin{figure*} 
   \includegraphics[width=0.7\textwidth,clip=]{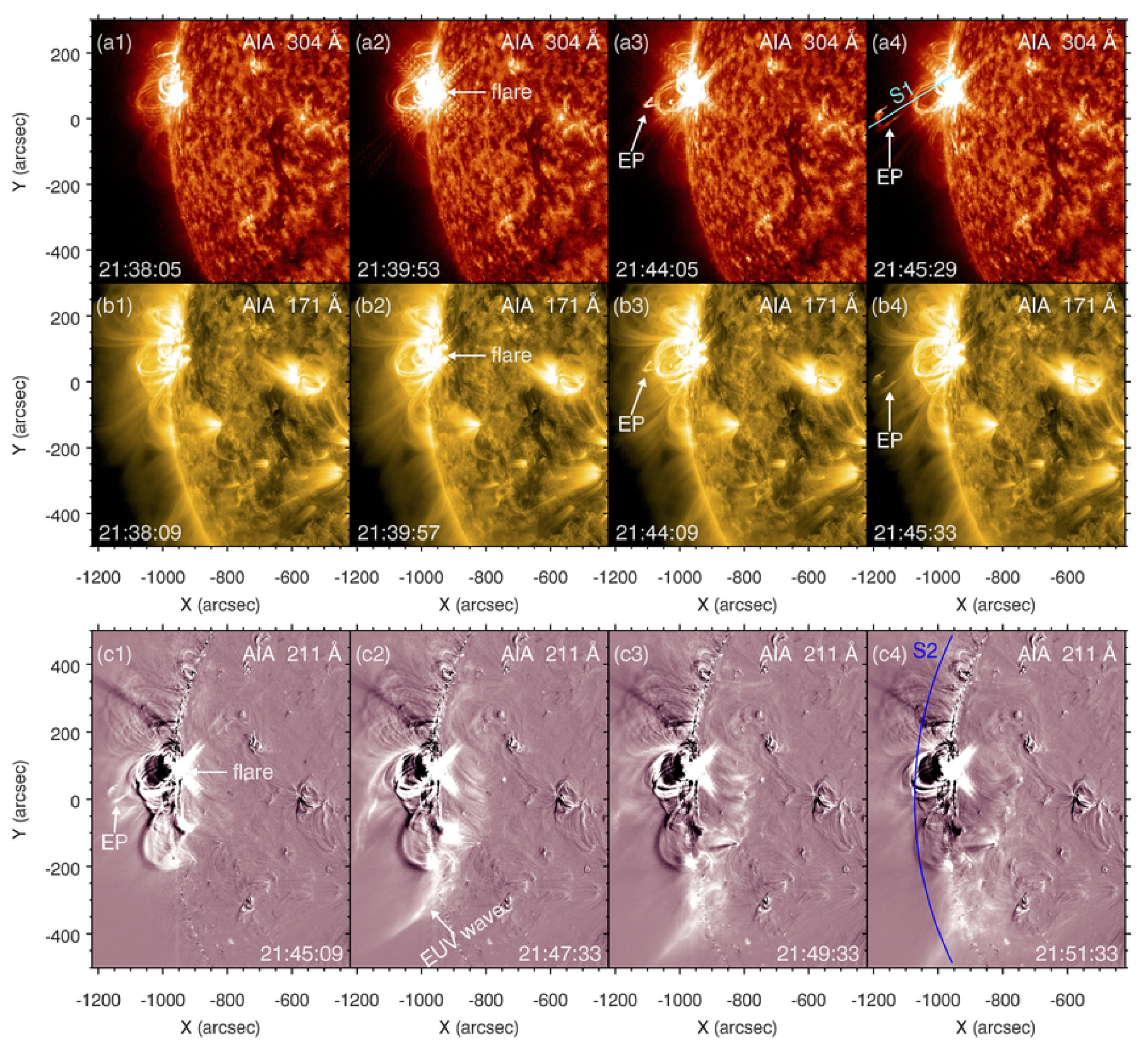}
   \centering
    \caption{Top and middle panels: snapshots of AIA 304 and 171 {\AA} images showing the non-radial prominence eruption and X5.0 flare.
    In panel (a4), a straight slice (S1) is used to investigate the evolution of prominence (EP).
    Bottom panels: snapshots of AIA 211 {\AA} images showing the EUV wave propagating southward.
    In panel (c4), a curved slice (S2) is used to investigate the evolution of EUV wave.}
    \label{fig1}
\end{figure*}

In Figure~\ref{fig1}, the top and middle panels show the eruptive prominence (EP) and X5.0 flare observed in SDO/AIA 304 and 171 {\AA} passbands.
The loop-like prominence is also evident in AIA 1600 {\AA} images \citep[see Fig. 2 in][]{cxy25}.
The EP originates from AR 13536 and rises up non-radially in the southeast direction.
The inclination angle between the EP and local vertical direction is $\sim$35$\fdg$4.
The bottom panels of Figure~\ref{fig1} show AIA 211 {\AA} base-difference images, 
featuring the EUV wave mainly propagating southward \citep[see Fig. 2 in][]{zxf25}.

\begin{figure*} 
	\includegraphics[width=0.7\textwidth,clip=]{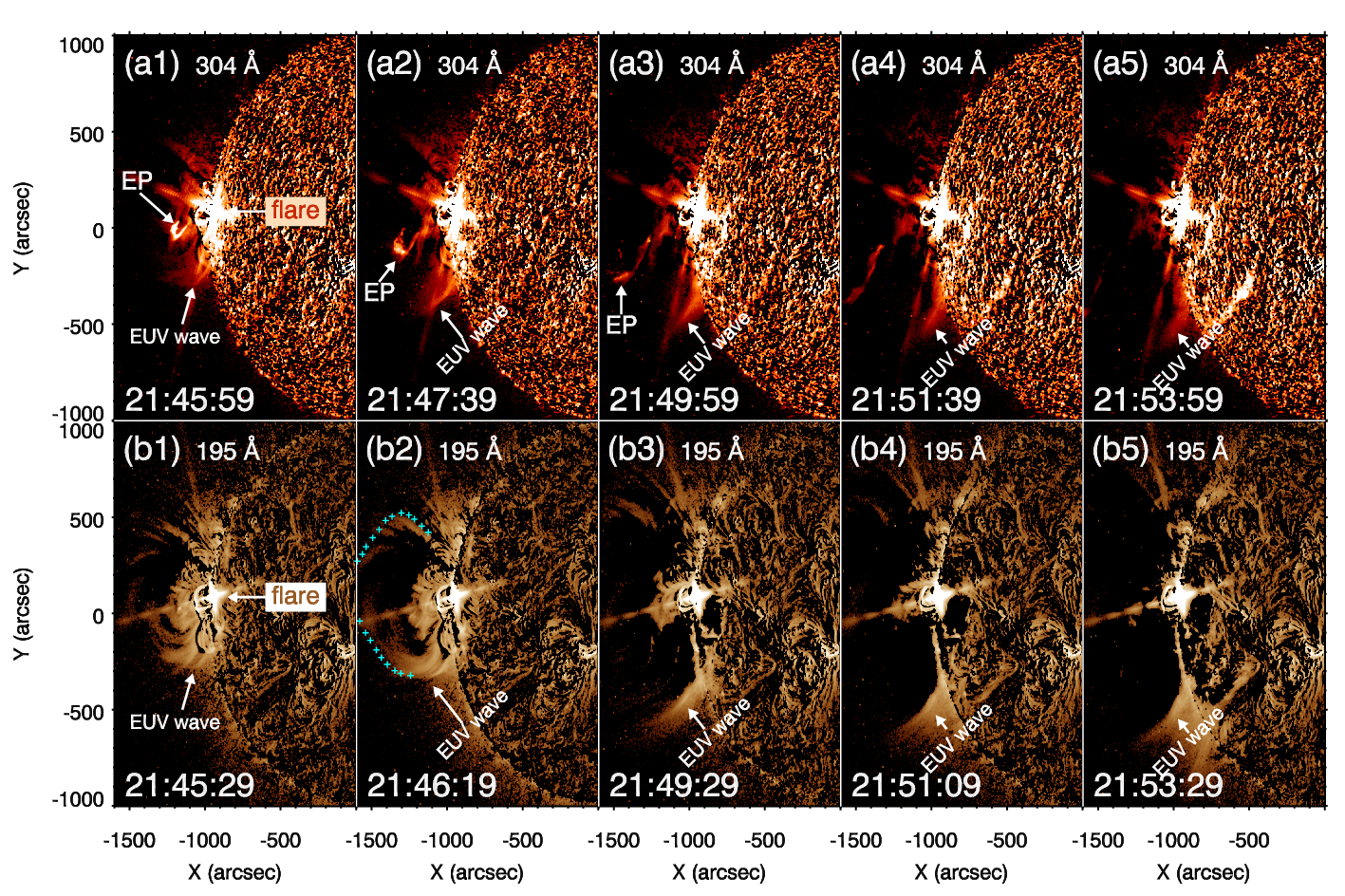}
	\centering
	\caption{GOES-16/SUVI base-difference images in 304 {\AA} (top panels) and 195 {\AA} (bottom panels).
	The EP, flare, and EUV wave are labeled.
	In panel (b2), the cyan plus symbols denote the CME leading edge.}
	\label{fig2}
\end{figure*}

\begin{figure} 
\includegraphics[width=0.45\textwidth]{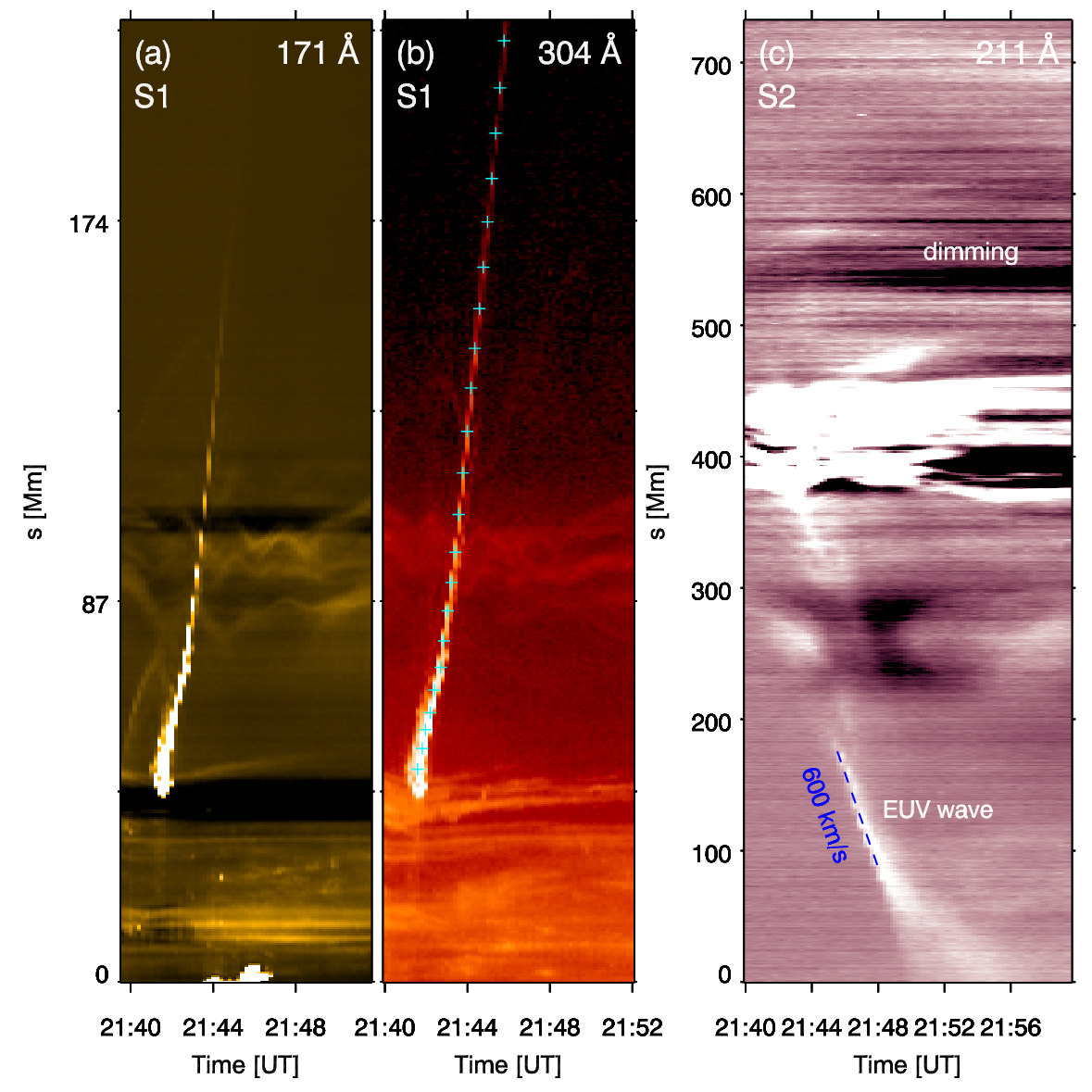}
\centering
\caption{(a)-(b) Time-distance diagrams of S1 in AIA 171 and 304 {\AA}.
              The cyan pluses signify the trajectory of the prominence along S1.
              (c) Time-distance diagram of S2 in AIA 211 {\AA}. The coronal dimming and speed of EUV wave ($\sim$600 km s$^{-1}$) along S2 are labeled.}
\label{fig3}
\end{figure}

In Figure~\ref{fig2}, the top and bottom panels show the EP, flare, and EUV wave in 304 and 195 {\AA} base-difference images observed by GOES-16/SUVI 
with a larger field of view (FOV) than SDO/AIA. The EP and EUV wave have the same morphology as those in AIA images.
The EUV wave starts at $\sim$21:45:30 UT and moves on for nearly $\sim$9 minutes \citep[see Fig. 3 in][]{zxf25}.
In panel (b2), the cyan plus symbols mark the CME leading edge observed at 21:46:19 UT when the shock and EUV wave form.
The top of leading edge is estimated to have a heliocentric distance of $\sim$1.75 $R_{\sun}$.
To investigate the kinematics of EP and EUV wave, we select a straight slice (S1) in Figure~\ref{fig1}(a4) and a curved slice (S2) in Figure~\ref{fig1}(c4) 
with total lengths of $\sim$303$\arcsec$ and $\sim$1011$\arcsec$, respectively.
Time-distance diagrams of S1 in AIA 171 and 304 {\AA} during 21:40$-$21:52 UT are displayed in Figure~\ref{fig3}(a)-(b).
Trajectory of the EP along S1 is denoted with cyan plus symbols.
Time-distance diagram of S2 in AIA 211 {\AA} during 21:40$-$22:00 UT is displayed in Figure~\ref{fig3}(c).
The speed of EUV wave along S2 is calculated to be $\sim$600 km s$^{-1}$.
To calculate the speed of EUV wave, \citet{zxf25} selected a different slice, which starts from the flare site and passes through a coronal hole.
Two components of the EUV wave are derived. The faster component at a speed of $\sim$948 km s$^{-1}$ is explained by a coronal shock wave.
It is noted that the two slices are situated at different altitudes. Therefore, a discrepancy in their measured velocities is reasonable \citep{shen12}.

\begin{figure*} 
\includegraphics[width=0.75\textwidth,clip=]{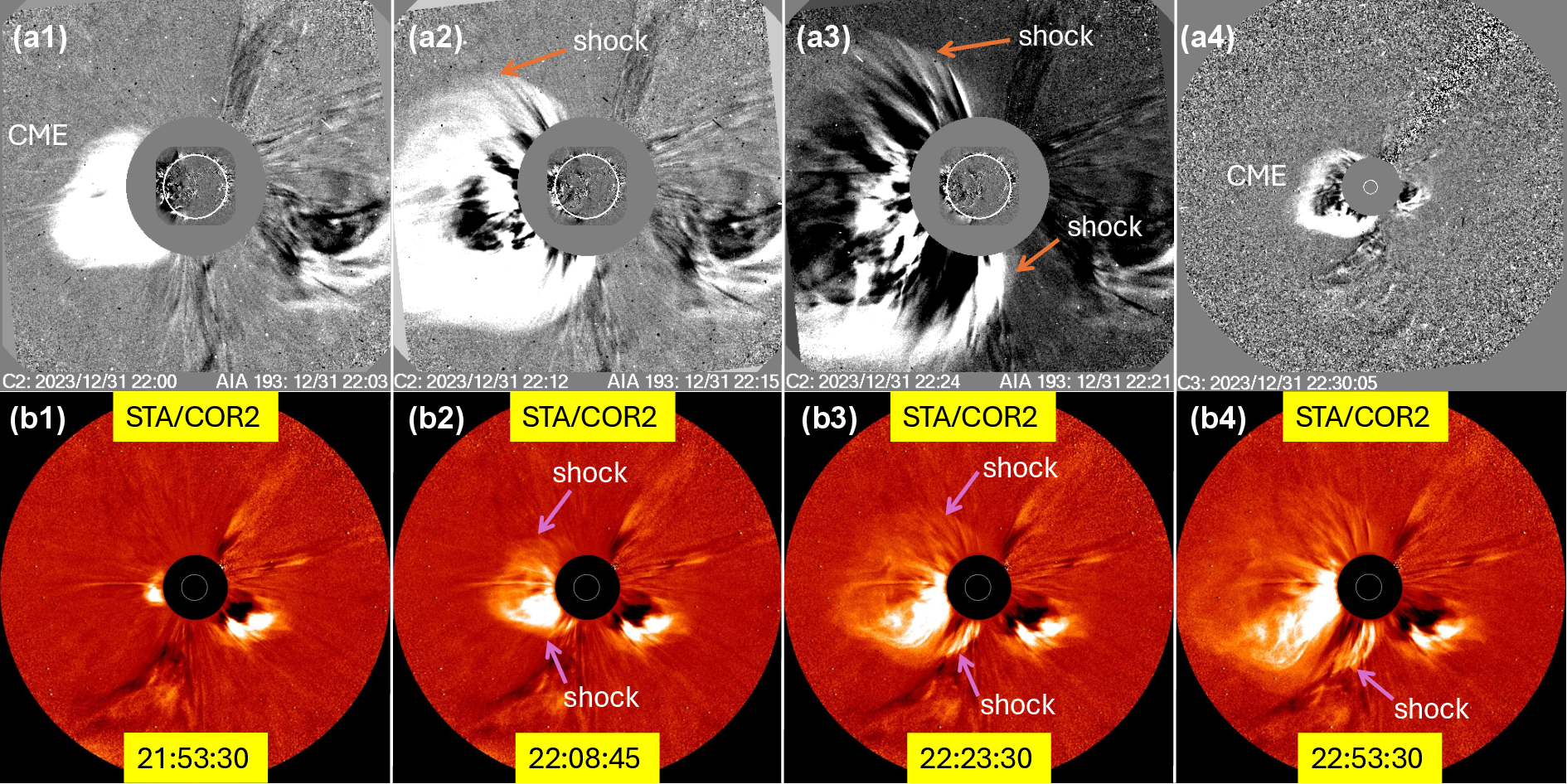}
\centering
\caption{WL images of the halo CME and shock observed by LASCO-C2 (a1)-(a3), LASCO-C3 (a4), and STA/COR2 (b1)-(b4) during 21:53$-$22:53 UT.}
	     \label{fig4}
\end{figure*}

\begin{figure} 
\includegraphics[width=0.55\textwidth]{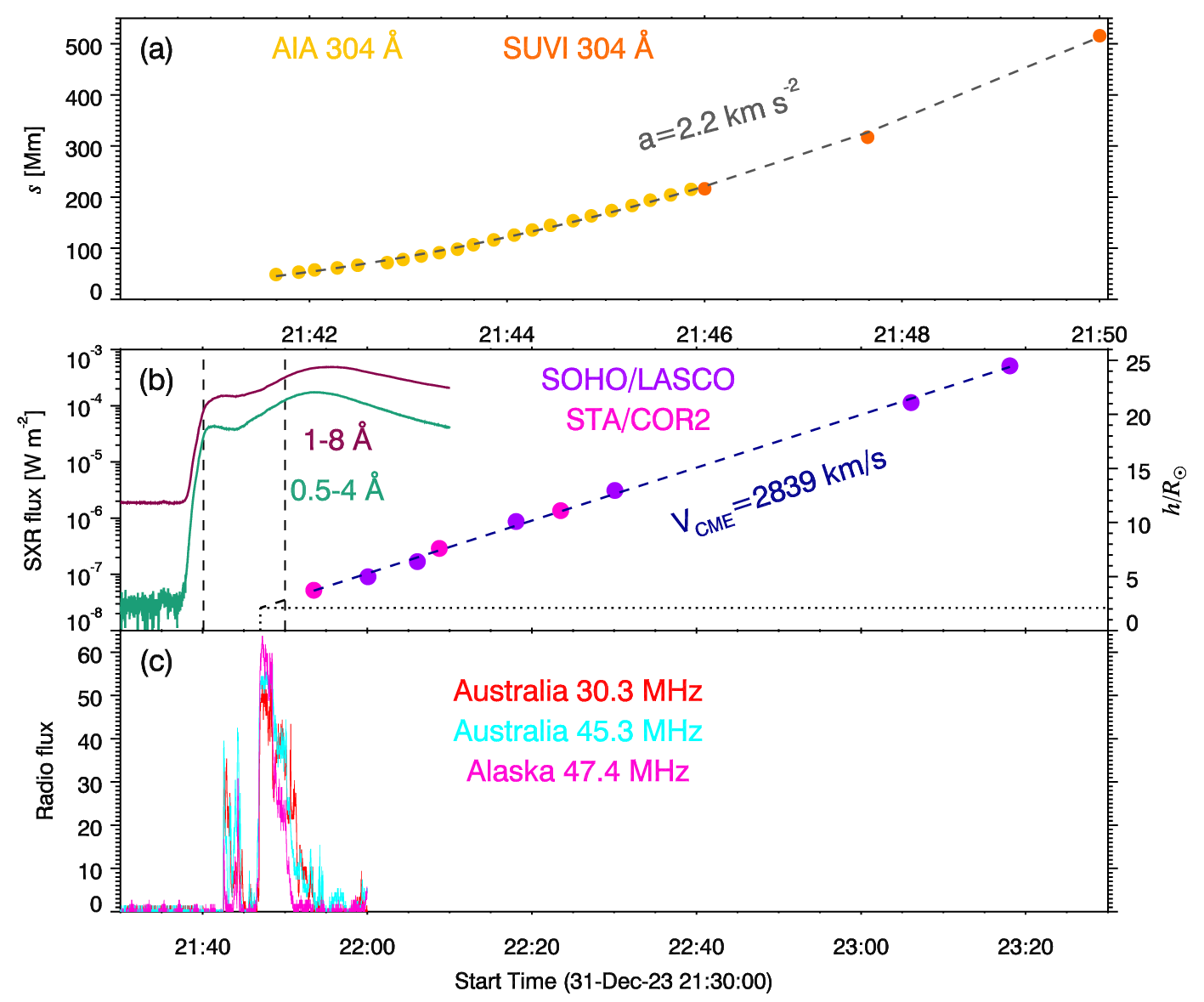}
\centering
\caption{(a) Trajectory of the EP along S1 in the FOV of AIA (yellow circles) and SUVI (orange circles).
              A curve fitting (grey line) using a quadratic function results in an acceleration of $\sim$2.2 km s$^{-2}$.
              (b) SXR light curves of the flare in 1$-$8 {\AA} (maroon line) and 0.5$-$4 {\AA} (green line).
              Two vertical dashed lines signify the time range of panel (a).
              Heights of the CME leading edge in the FOVs of SOHO/LASCO and STA/COR2 are drawn with purple and magenta circles.
              A linear fitting (dark blue line) results in an apparent speed of $\sim$2839 km s$^{-1}$.
              (c) Radio fluxes of the herringbones at 30.3, 45.3, and 47.4 MHz.}
\label{fig5}
\end{figure}

In Figure~\ref{fig4}, the top and bottom panels show the halo CME and shock observed by SOHO/LASCO and STA/COR2 during 21:53$-$22:53 UT.
The CME first appears at 21:53:30 UT and 22:00:05 UT in the FOVs of COR2 and LASCO with a central position angle of $\sim$112$\degr$.
As it moves forward, a shock is driven ahead and at the CME flank, which are pointed by orange and magenta arrows.
In Figure~\ref{fig5}(b), time evolutions of the CME leading edge height in the FOVs of LASCO and COR2 are plotted with purple and magenta circles, respectively.
A linear fitting (blue line) results in an apparent speed of $\sim$2839 km s$^{-1}$.
Combining the observations of AIA, SUVI, and LASCO, \citet{cxy25} concluded that the CME leading edge also experiences a considerable acceleration.
The speed increases rapidly from $\sim$1025 km s$^{-1}$ during 21:43:49$-$21:47:49 UT to $\sim$2807 km s$^{-1}$ during 22:00:05$-$22:30:05 UT.
Although there is a data gap of CME observation between 21:48$-$21:53 UT, 
it is inferred that the CME leading edge experiences a continuous acceleration between 21:43 UT and 22:00 UT.

In Figure~\ref{fig5}(a), the distances of EP along the direction of S1 in the FOVs of AIA and SUVI 304 {\AA} are drawn with yellow and orange circles, respectively.
The height of EP increases rapidly from $\sim$48.5 to $\sim$515.8 Mm within $\sim$500 s.
A quadratic fitting (gray dashed line) indicates a significant acceleration of $\sim$2.2 km s$^{-1}$.
The speed of EP increases from $\sim$393.5 to $\sim$1476.4 km s$^{-1}$.
In Figure~\ref{fig5}(b), SXR light curves of the flare during 21:30$-$22:10 UT in 1$-$8 {\AA} and 0.5$-$4 {\AA} are drawn with maroon and green lines, respectively.
The SXR fluxes start to increase at $\sim$21:36 UT, peak at $\sim$21:55 UT before declining gradually \citep[see Fig. 1 in][]{ryan24}.
Hence, the fast acceleration of the prominence occurs in the impulsive phase of the flare \citep{cx20}.
Similarly, during the non-radial propagation on 2011 February 24, the prominence experiences a remarkable acceleration ($\sim$0.98 km s$^{-2}$),
which is coincident with the impulsive phase of the related flare \citep{lsy25}.
In our case, the acceleration of prominence is $\sim$2.2 times higher than that of \citet{lsy25}.
 
\begin{figure} 
		\includegraphics[width=0.75\textwidth,clip=]{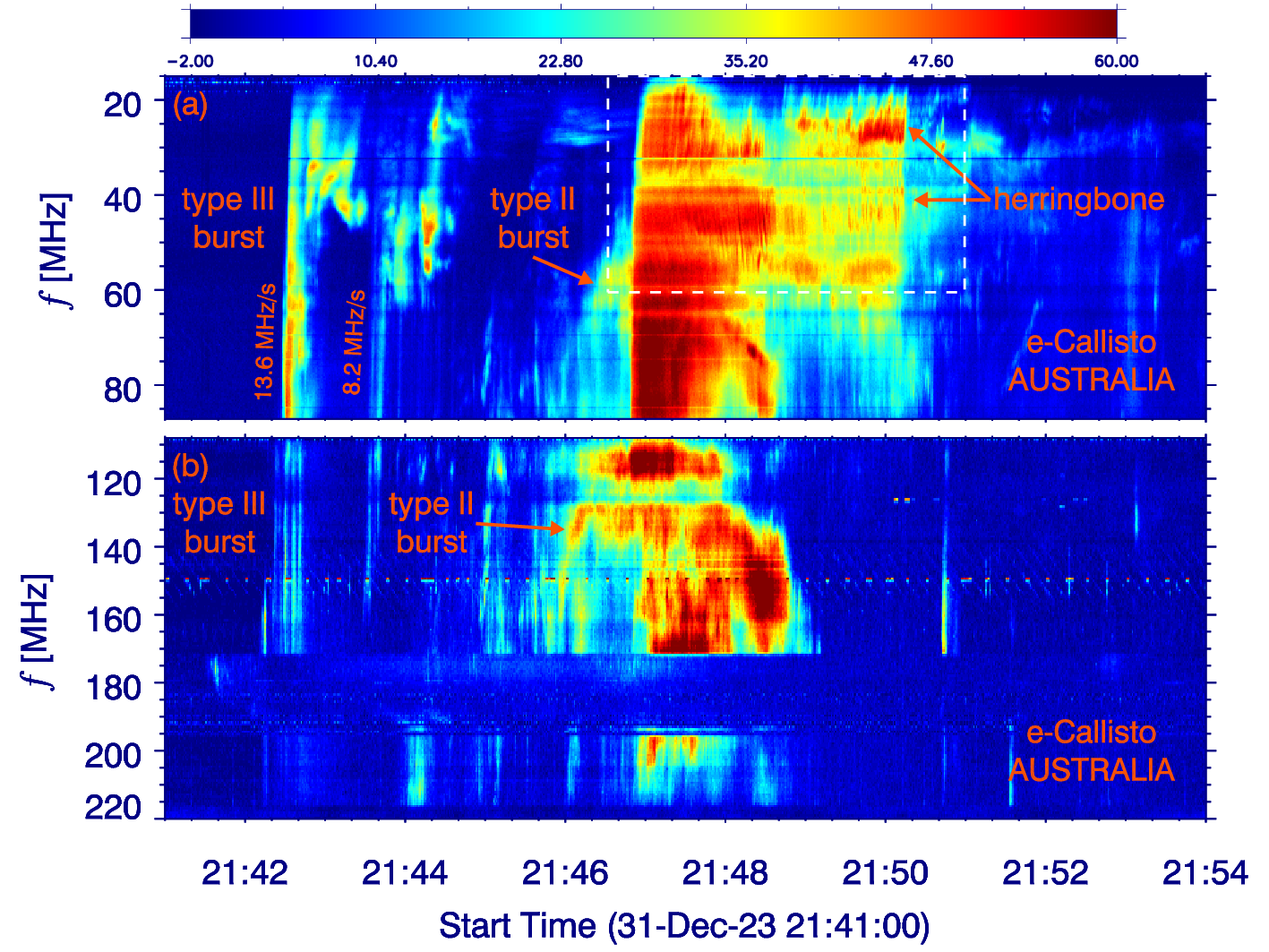}
		\centering
		\caption{Radio dynamic spectra of the flare observed by e-Callisto AUSTRALIA 
		in the frequency range 15$-$87 MHz (top panel) and 108$-$220 MHz (bottom panel).
		The type III bursts, type II burst, and herringbone structures are labeled.
		The dynamic spectra enclosed by the white dashed box are drawn in Figure~\ref{fig9}.}
		\label{fig6}
	\end{figure}

Figure~\ref{fig6} shows radio dynamic spectra of the flare observed by e-Callisto AUSTRALIA 
in the frequency range 15$-$87 MHz (top panel) and 108$-$220 MHz (bottom panel) during 21:41$-$21:54 UT.
A type III burst could be clearly identified in the frequency range 20$-$170 MHz during 21:42:20$-$21:43:00 UT.
The second but much weaker type III burst occurs around 21:43:40 UT.
It is generally accepted that type III bursts are produced by plasma emission of high-energy electrons 
escaping from the flare site along open magnetic field \citep{asch95,cla21}.
Two red arrows point to the type II burst starting from $\sim$21:45:40 UT, with the frequency drifting from $\sim$160 MHz to lower values gradually.
It is noticed that the beginning time of type II burst is coincident with that of EUV wave (Figure~\ref{fig2}), 
indicating that the EUV wave and CME-driven shock are launched simultaneously.
The subregion within the white dashed box features herringbone structures superposed on the type II burst, which will be described in detail in the next Section.

\begin{figure} 
   \includegraphics[width=0.30\textwidth,clip=]{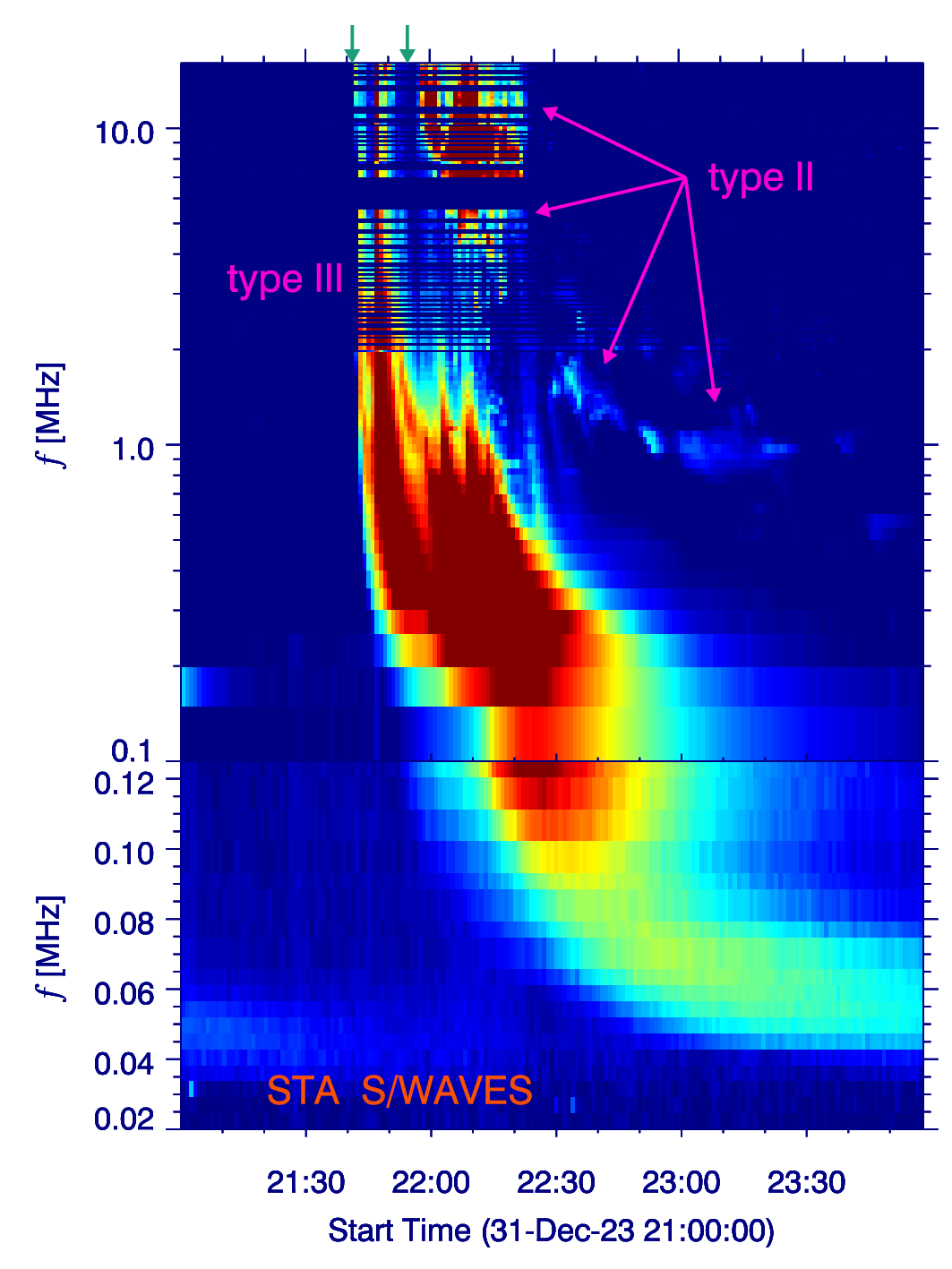}
   \centering
   \caption{Radio dynamics spectra observed by S/WAVES on board STA during 21:00$-$23:59 UT, showing the type III and type II bursts.
                 Two green arrows point to the start and end moments corresponding to Figure~\ref{fig6}.}
   \label{fig7}
\end{figure}

The type III and type II bursts were also detected by S/WAVES on board STA with a lower cadence.
Figure~\ref{fig7} shows the radio dynamic spectra of S/WAVES during 21:00$-$23:59 UT.
Two green arrows point to the start (21:41 UT) and end (21:54 UT) moments of Figure~\ref{fig6}.
The frequency of type III bursts decreases from $\sim$16 MHz to $\sim$40 kHz \citep[see Fig. 1 in][]{cxy25}.
The magenta arrows point to the type II burst, whose frequency decreases from $\sim$16 MHz to $\sim$1 MHz.
Considering the time and frequency ranges, it could be correlated with the propagation of CME and associated shock in Figure~\ref{fig4}
and considered as a natural extension of type II burst displayed in Figure~\ref{fig6}.
	
\subsection{Herringbone structures} \label{hb}
\begin{figure}
\includegraphics[width=0.70\textwidth,clip=]{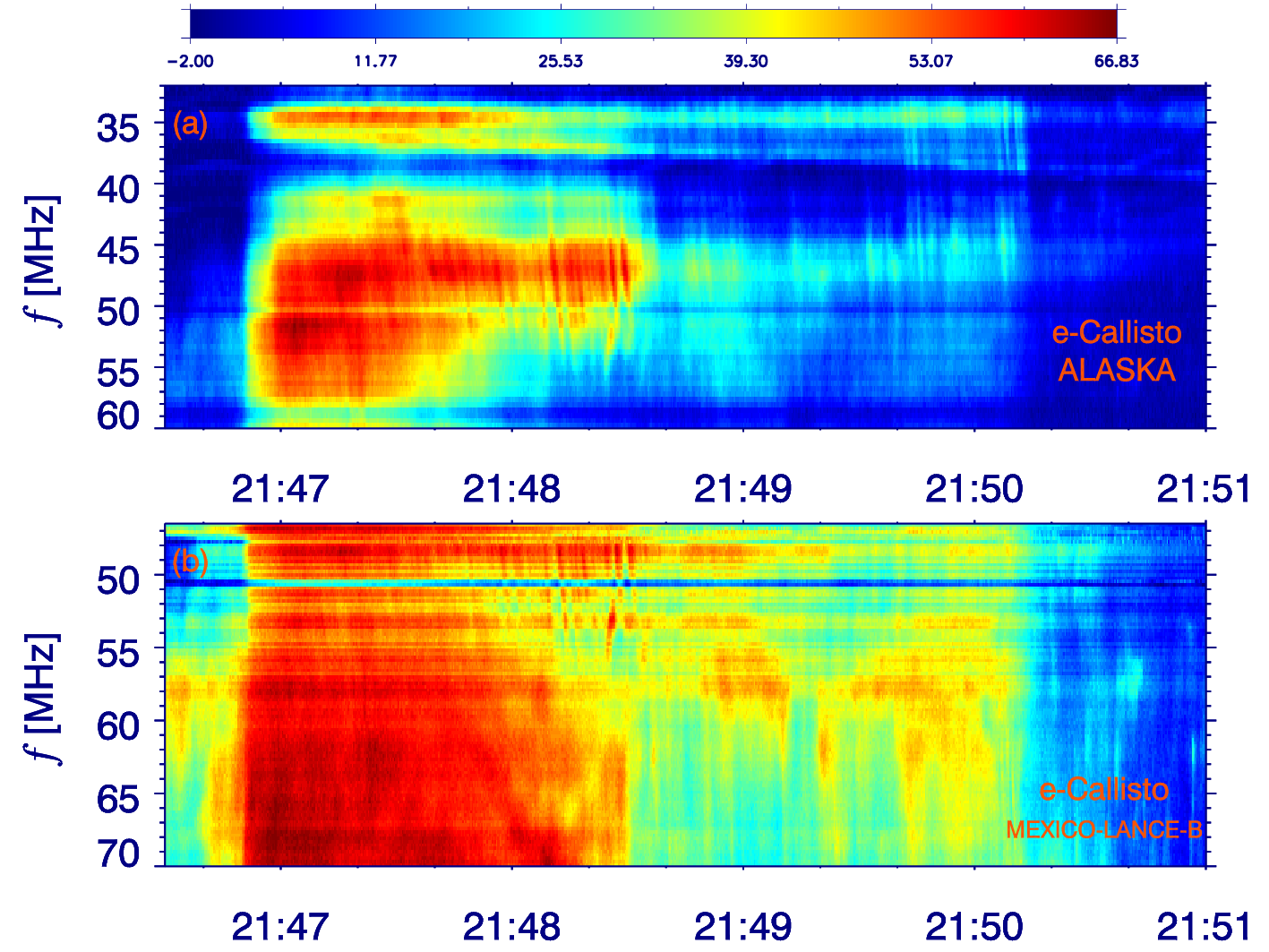}
\centering
\caption{Radio dynamic spectra of the flare observed by e-Callisto ALASKA (a) and MEXICO-LANCE-B (b),
              featuring reverse-drift herringbone structures.}
\label{fig8}
\end{figure}

As shown in Figure~\ref{fig6}(a), there are simultaneous forward-drift and reverse-drift structures during 21:47$-$21:51 UT, i.e., herringbone structures \citep{car13,mor19}.
Figure~\ref{fig8} shows the radio dynamic spectra observed by e-Callisto ALASKA (top panel) and MEXICO-LANCE-B (bottom panel) during 21:46:30$-$21:51:00 UT.
Due to the limitations of frequency range, only reverse-drift structures of the herringbones are detected by these telescopes, while the forward-drift structures are missing.

\begin{figure} 
		\includegraphics[width=0.60\textwidth,clip=]{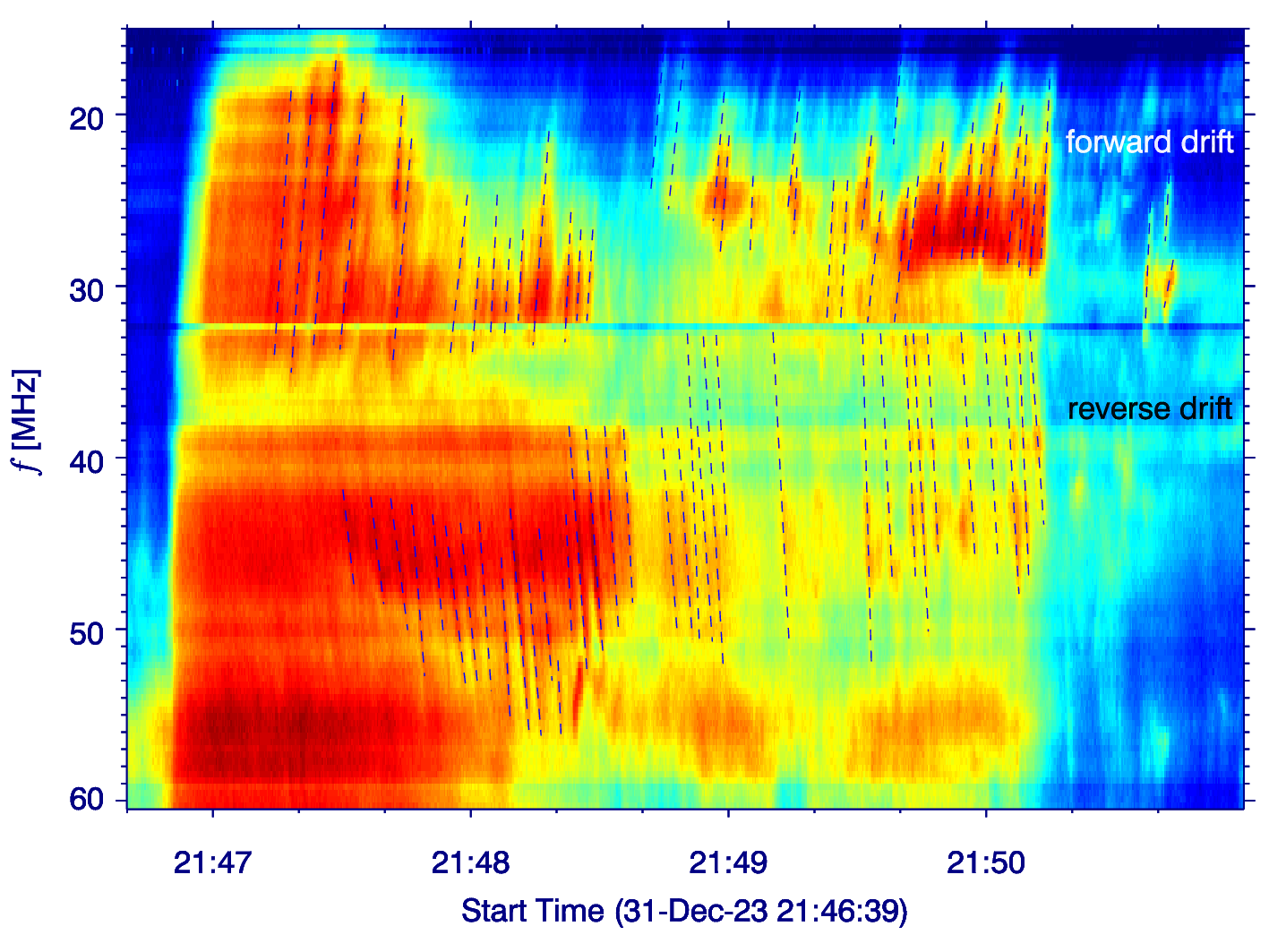}
		\centering
		\caption{A close-up of the dynamic spectra during 21:46:39$-$21:51:00 UT.
		Manually identified forward-drift and reverse-drift bursts are superposed with blue dashed lines.
		The slopes of these bursts are used to calculate the frequency drift rates.}
		\label{fig9}
	\end{figure}

\citet{car13} investigated the herringbones associated with the CME on 2011 September 22. 
In a follow-up work, using the image processing technique called Hough Transform \citep{hou61}, 
\citet{car15} identified 188 herringbone fine structures automatically to perform a statistical analysis.
The drift rates, electron velocities, and beam displacements are derived and investigated in detail.
Figure~\ref{fig9} shows a close-up of the dynamic spectra enclosed by the white dashed box in Figure~\ref{fig6}(a).
The time range is between 21:46:39 UT and 21:51:00 UT. The frequency range is from 15 to 60 MHz.
In order to explore the herringbones in the current study, we identified 40 forward-drift and 35 reverse-drift bursts manually, 
which are superposed with blue dashed lines \citep[similar to Fig. 2 in][]{car15}.
Parameters of these bursts could be readily obtained, 
including the duration ($\Delta t$), band width (BW), starting frequency ($f_{s}$), frequency drift rate ($D_{f}=\frac{df}{dt}$), and maximum intensity ($I_{max}$).

To estimate the velocities of electron beams ($v$) producing the herringbones, we use the following formula \citep{mor19}:
\begin{equation} \label{eqn-1}
    v=\frac{2n_{e}D_{f}}{f}(\frac{dn_{e}}{dr})^{-1},        
\end{equation}
where $n_{e}(r)$ denotes the electron number density as a function of heliocentric distance ($r$ in units of $R_{\sun}$).
$\frac{dn_{e}}{dr}$ represents the density gradient.
Here, we adopt the Newkirk electron density model \citep{new61}, which has been widely used:
\begin{equation} \label{eqn-2}
    n_{e}(r)=4.2\times10^{4}\times10^{4.32/r}.
\end{equation}
The frequency of plasma emission $f(r)$ depends solely on the electron number density $n_{e}(r)$:
\begin{equation} \label{eqn-3}
    f(r)=8.98\times \sqrt{n_{e}(r)}.
\end{equation}

\begin{figure} 
		\includegraphics[width=0.55\textwidth,clip=]{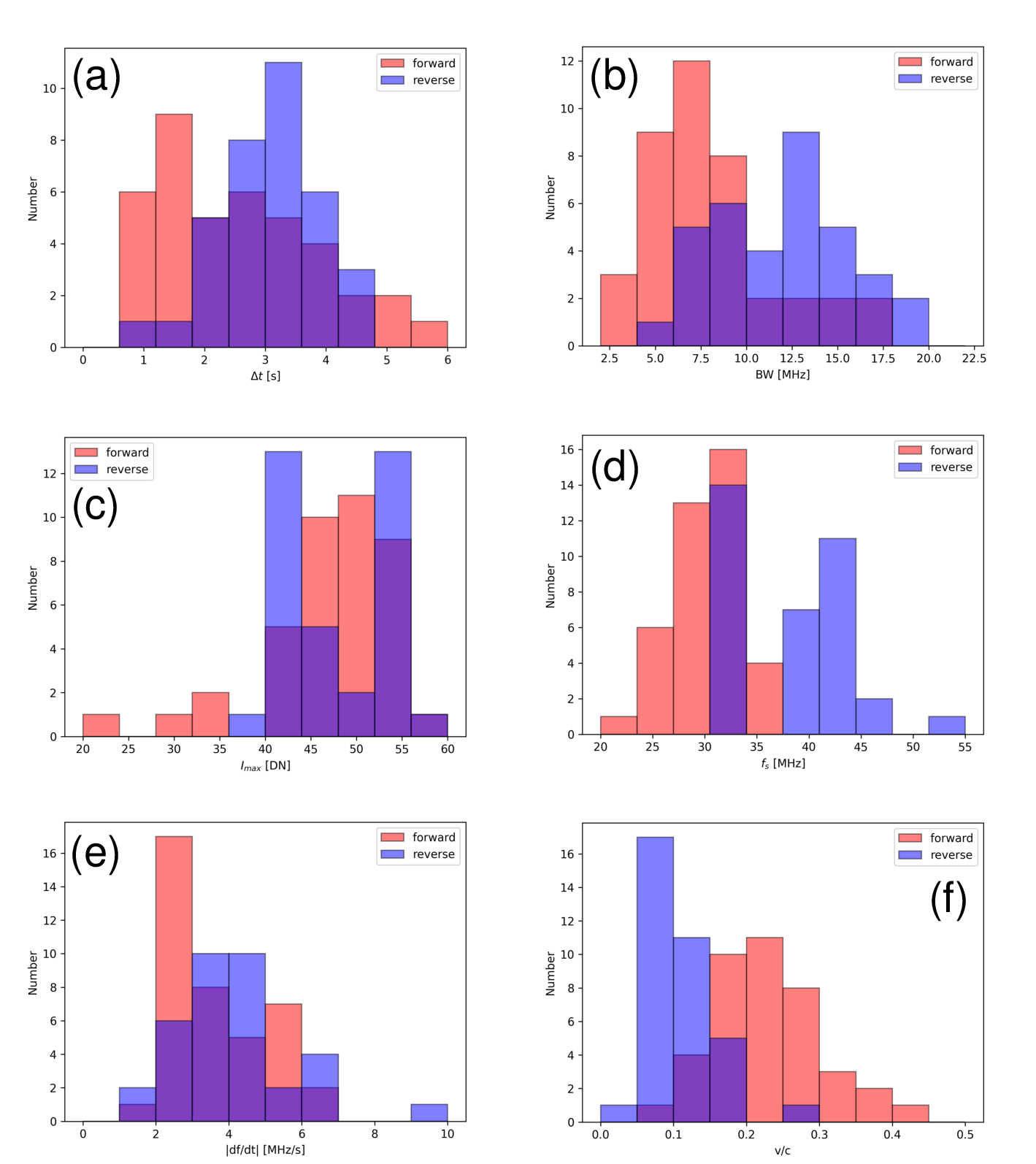}
		\centering
		\caption{Histograms of the parameters for forward-drift (red color) and reverse-drift (blue color) herringbone bursts,
		including the duration (a), band width (b), maximum intensity (c), starting frequency (d), frequency drift rate (e), and electron beam velocity (f).}
		\label{fig10}
	\end{figure}

Figure~\ref{fig10} shows the histograms of these parameters for forward-drift (red color) and reverse-drift (blue color) bursts.
The minimum, maximum, and mean values are listed in Table~\ref{tab2}. 
In panel (a), the duration of these fine structures are between $\sim$0.7 and $\sim$5.6 s, with mean values of 2.5$-$3.1 s,
which are very close to the statistical results of \citet{car15}.
In panel (b), the band widths are between $\sim$2.8 and $\sim$20 MHz. The average band width for reverse-drift bursts is $\sim$1.5 times higher than that of forward-drift ones.
In panel (e), the absolute drift rates are between $\sim$1.3 and $\sim$9.4 MHz s$^{-1}$, with mean values of 3.6$-$4.1 MHz s$^{-1}$, 
which account for one third to half of the drift rates of type III bursts in Figure~\ref{fig6}(a).
\citet{doro15} carried out a statistical analysis on forward-drift and reverse-drift bursts of herringbone structures in the frequency range 12$-$30 MHz. 
The average values of $|\frac{df}{dt}|$ are $\sim$1.23 and $\sim$1.80 MHz s$^{-1}$, which account for one third to half of the values in our sample.
The average values of duration are $\sim$2.3 s and $\sim$1.9 s, which are slightly lower than our results.

In panel (f), the corresponding electron velocities derived using Equation~\ref{eqn-1} fall in the range of 0.04$-$0.41 $c$.
The mean value of $v$ for forward-drift bursts is almost twice higher than that of reverse-drift bursts.
On the contrary, \citet{car15} found that the speed distributions of beams producing forward and reverse drift bursts show a slight difference 
and the total histogram peaks at $\sim$0.16 $c$.
According to Equation~\ref{eqn-1}, the beam velocity is inversely proportional to the frequency and density gradient.
As shown in Figure~\ref{fig10}(d)-(e), the drift rates of forward-drift and reverse-drift bursts are comparable, 
while the starting frequencies have notable differences. Moreover, the density gradients of reverse-drift bursts are larger than those of forward-drift bursts.
Therefore, the average velocity of beams are higher in forward-drift bursts. Additional statistical works are required to confirm the conjecture.
It should be emphasized that the results of $D_{f}$ and $v$ may vary when adopting different density models \citep[e.g.,][]{sai77,leb98,mann99}.

In panel (c), the maximum intensities of these bursts lie in the range of 23$-$54 DN. The average values of forward-drift and reverse-drift bursts are almost equal.
In panel (d), the starting frequency ($f_{s}$) lies in the range of 21$-$52 MHz, with mean values of 30$-$38 MHz.
According to Equation~\ref{eqn-3}, the heights of shock acceleration regions are 0.64$-$0.78 $R_{\sun}$ (448$-$546 Mm) above the photosphere,
which are basically in agreement with the height of CME leading edge when the shock forms (Figure~\ref{fig2}(b2)).
In the event on 2011 September 22, the separatrix frequency between the forward-drift and reverse-drift structures drifts from $\sim$43 to $\sim$32 MHz \citep{car15}.
The corresponding height of shock acceleration increases from $\sim$0.57 to $\sim$0.74 $R_{\sun}$ using the Newkirk model.
In the event on 2017 September 10, the separatrix frequency is around 40 MHz.
It is noted that the fine structures of herringbone in our study are selected by eye rather than automatically. 
Consequently, the derived parameters of these bursts may have some uncertainties.
Nevertheless, comparisons with previous works indicate that the average values are generally in accordance with their findings.
	
\begin{figure} 
		\includegraphics[width=0.45\textwidth,clip=]{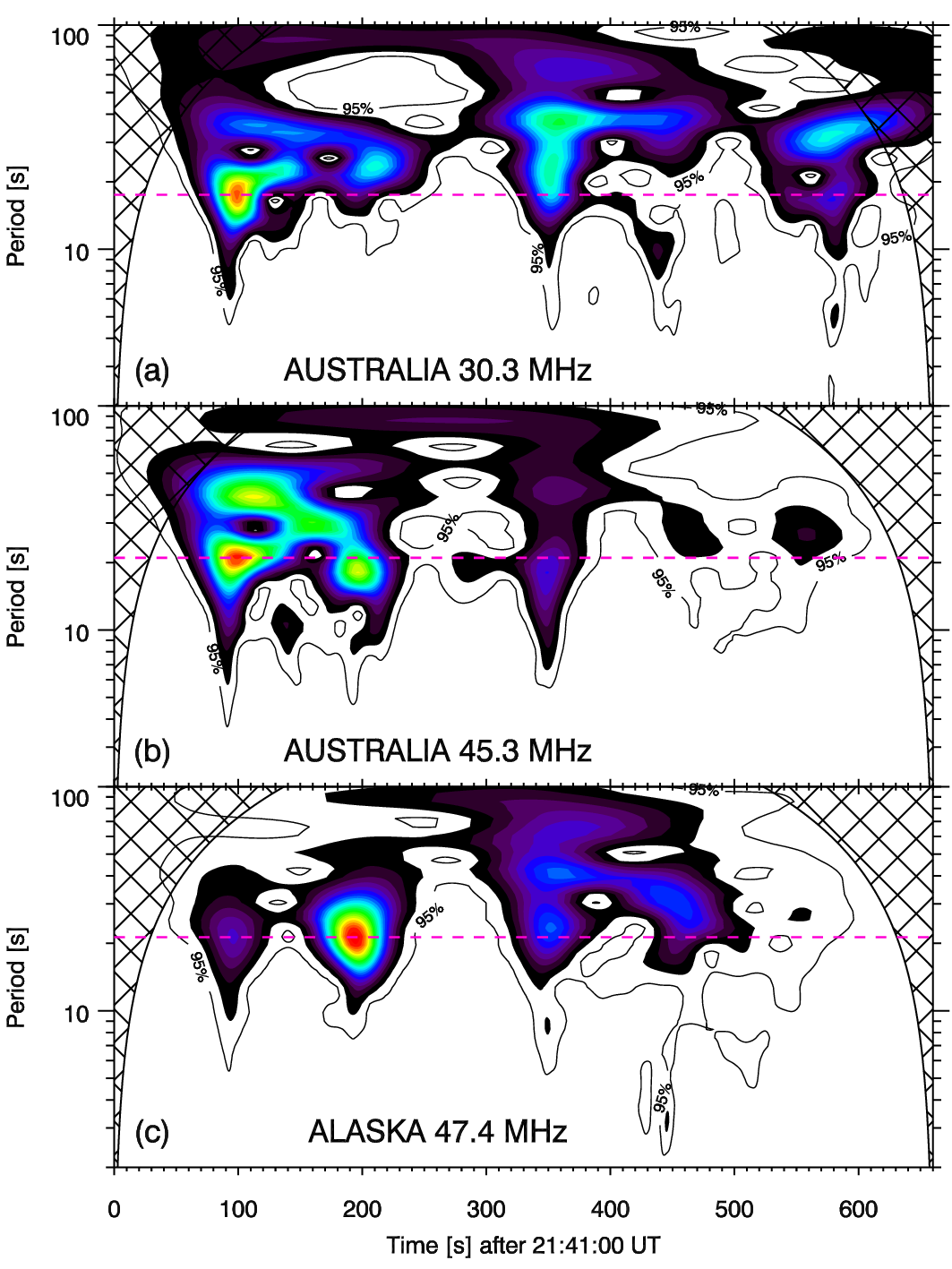}
		\centering
		\caption{Wavelet transforms of the radio fluxes at 30.3 (a), 45.3 (b), and 47.4 MHz (c).
		The magenta dashed lines represent the corresponding periods.}
		\label{fig11}
	\end{figure}

The radio dynamic spectra of AUSTRALIA, ALASKA, and MEXICO-LANCE-B (Figures~\ref{fig8} and \ref{fig9}) 
show that the bursts of herringbone structures appear intermittently,
suggesting that electrons are accelerated quasi-periodically \citep{car13}. 
To explore the periods of herringbones, we extract the radio fluxes at 30.3 and 45.3 MHz from AUSTRALIA observations and 47.4 MHz from ALASKA observations.
Time evolutions of these fluxes are plotted in Figure~\ref{fig5}(c). 
The slowly-varying components of the fluxes are derived using the smoothing method with a window of $\sim$30 s.
The fast-varying components are background-removed fluxes. The wavelet transforms of the fast-varying components are displayed in Figure~\ref{fig11}.
It is clear that the radio fluxes have QPPs with periods of $\sim$17.5, $\sim$21.0, and $\sim$21.3 s at 30.3, 45.3, and 47.4 MHz.
These periods are longer than those reported by \citet{car13}.

\begin{deluxetable*}{c|c|cccccc}
		\digitalasset
		\tablewidth{\textwidth}
		\tablecaption{Statistic analysis of forward-drift and reverse-drift herringbone structures, 
		including the duration ($\Delta t$), band width (BW), frequency drift rate ($\frac{df}{dt}$), starting frequency ($f_s$), 
		maximal intensity ($I_{max}$), and electron velocity ($v$).
		\label{tab2}}
		\tablecolumns{8}
		\tablenum{2}
		\tablehead{
		        \colhead{Type} &
			\colhead{Value} &
			\colhead{$\Delta t$} &
			\colhead{BW} &
			\colhead{$|\frac{df}{dt}|$} &
			\colhead{$f_{s}$} &
			\colhead{$I_{max}$} &
			\colhead{$v$} \\
			\colhead{} &
			\colhead{} &
			\colhead{(s)} &
			\colhead{(MHz)} &
			\colhead{(MHz s$^{-1}$)} &
			\colhead{(MHz)} &
			\colhead{(DN)} &
			\colhead{($c$)}
		}
		\startdata
                 Forward-drift & Min.    & 0.69 & 2.78   & 1.33 & 21.74 & 23.58 & 0.07 \\
                 Forward-drift & Max.   & 5.56 & 16.94 & 6.39 & 35.05 & 56.83 & 0.41 \\
                 Forward-drift & Mean  & 2.54 & 8.12   & 3.61 & 30.18 & 47.38 & 0.23 \\
                 \hline
                 Reverse-drift & Min.    &   0.69 & 4.29 & 1.95 & 32.61 & 38.40 & 0.04 \\
                 Reverse-drift & Max.   &  4.64 & 19.64 & 9.44 & 51.82 & 56.36 & 0.26 \\
                 Reverse-drift & Mean  &  3.07 & 11.91 & 4.13 & 38.45 & 48.15 & 0.11 \\
		\enddata
\end{deluxetable*}

\section{Summary and discussions} \label{sum}
In this paper, we report quasi-periodic herringbone structures during the impulsive phase of an X5.0 flare, 
coinciding with the distinct acceleration phase of eruptive prominence ejection.
The flare results from the prominence eruption in NOAA AR 13536 on 2023 December 31.
The prominence propagates non-radially in the southeast direction with an inclination angle of $\sim$35$\fdg$4.
The speed of prominence increases from $\sim$394 to $\sim$1476 km s$^{-1}$ within $\sim$500 s.
The fast CME at a speed of $\sim$2852 km s$^{-1}$ drives a shock wave and a coronal EUV wave observed in WL coronagraphs and EUV images, respectively.
The herringbone structures lasting for $\sim$4 minutes take place at the initial stage of a group of type II burst.
The herringbones in the frequency range 20$-$70 MHz are characterized by simultaneous forward-drift and reverse-drift stripes 
with average durations of $\sim$2.5 s and $\sim$3.1 s.
The frequency drift rates of these bursts range from $\sim$1.3 to $\sim$9.4 MHz s$^{-1}$
with average values of $\sim$3.6 and $\sim$4.1 MHz s$^{-1}$, respectively. 
The heights of particle acceleration regions are estimated to be 0.64$-$0.78 $R_{\sun}$ (448$-$546 Mm) above the photosphere.
The speeds of electron beams producing the herringbones are estimated to be 0.04$-$0.41 $c$, 
with average values of $\sim$0.23 $c$ and $\sim$0.11 $c$ for forward-drift and reverse-drift bursts.
QPPs with periods of 17.5$-$21.3 s are discovered in the radio fluxes of herringbones, indicating quasi-periodic accelerations of electrons by the shock wave.

	\begin{figure} 
		\includegraphics[width=0.75\textwidth,clip=]{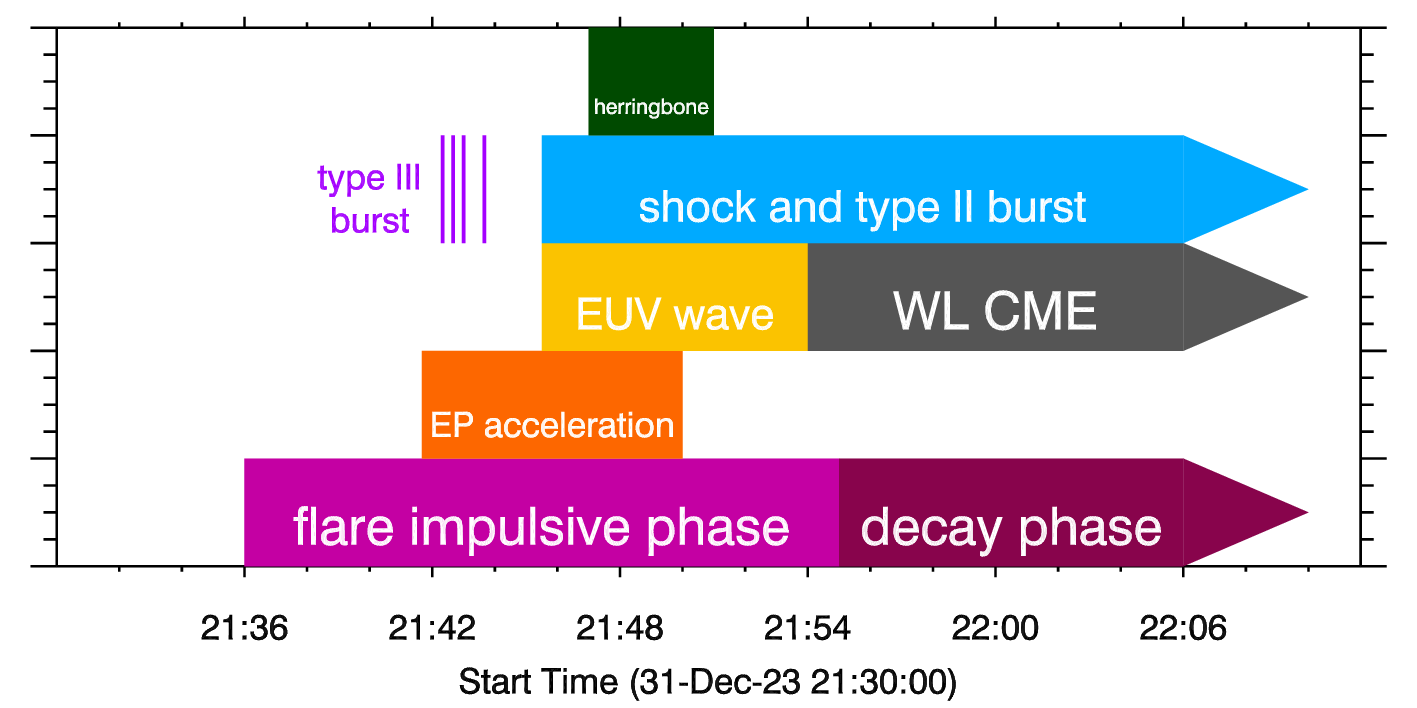}
		\centering
		\caption{Time line of the activities during the eruptive flare on 2023 December 31.}
		\label{fig12}
	\end{figure}

Although shock waves and type II radio bursts associated with CMEs are frequently observed \citep{vrs06,gop13,gop19,zqm22,pan25}, 
herringbones structures are rarely detected \citep{cai87,mag20,car21,mor22,mor24}. In the era of SDO since 2010,
there are several eruptive flares showing unambiguous herringbone structures in the radio dynamic spectra \citep[e.g.,][]{car13,mor19}.
In Table~\ref{tab3}, the parameters of these events and the event in this study are summarized, 
including the date, location, flare class, CME speed ($v_{\mathrm{CME}}$), EUV wave speed ($v_{wave}$), Alfv\'{e}n Mach number ($M_{a}$) of the shock,
duration ($\tau_{\mathrm{HB}}$) and frequency range of herringbone ($f_{\mathrm{HB}}$), and electron beam speed ($v$).
All these events are related to shocks with large $M_{a}$.
It seems that herringbones are closely related to large eruptive flares, accompanied by strong shock waves, especially quasi-perpendicular shocks,
which are capable of accelerating electrons to high energies efficiently \citep{hol83,gia05,zank06,car21,mann22}.
To investigate how solar energetic particle events are created by coronal shocks, 
\citet{kong17} studied the process in which a fast shock propagates through a streamer-like magnetic field including both closed and open field lines.
Particles are efficiently accelerated due to perpendicular shock geometry.
\citet{mor22} investigated the herringbones associated with a coronal EUV wave generated by fast CME expansion on 2013 October 25.
It is concluded that the CME and related shock wave propagate quasi-perpendicularly to the streamer where herringbones are created.
The interaction between the CME shock and streamer is favorable to the production of electron beams that generate herringbones due to shock drift acceleration.
\citet{kov23} analyzed the type II burst associated with the CME/shock on 2014 July 25.
The herringbone type II was generated when the CME/shock penetrated into the lower part of the streamer below its cusp point.

Figure~\ref{fig12} shows a time line of the activities during the X5.0-class eruptive flare.
It is clear that the prominence acceleration occurs during the impulsive phase of the flare, giving rise to the formation of CME leading edge seen in EUV wavelengths.
The EUV wave and shock wave associated with the type II burst start at $\sim$21:45:30 UT.
The herringbone structures superposed on the type II burst are evident during 21:47$-$21:51 UT.
According to the schematic cartoons showing the locations of herringbone sources from different perspectives \citep[see Fig. 3 in][]{mor19},
the forward-drift and reverse-drift stripes are due to electron beams moving outward and toward the sun along open field.
Therefore, the high-energy electrons are released and herringbones are generated once the shock wave encounters open field during its expansion.
In the future, statistical analysis is worthwhile to figure out the physical origin and properties of herringbone structures in depth.
Numerical simulations are required to investigate particle accelerations by coronal shock waves.

\begin{deluxetable*}{cccccccccc}
		\digitalasset
		\tablewidth{\textwidth}
		\tablecaption{Comparison of the events producing herringbone structures.
		\label{tab3}}
		\tablecolumns{10}
		\tablenum{3}
		\tablehead{
			\colhead{Date} &
			\colhead{AR} &
			\colhead{Flare} &
			\colhead{$v_{\mathrm{CME}}$} &
			\colhead{$v_{wave}$} &
			\colhead{$M_{a}$} &
			\colhead{$\tau_{\mathrm{HB}}$} &
			\colhead{$f_{\mathrm{HB}}$} &
			\colhead{$v$} &
			\colhead{Reference} \\
			\colhead{ } &
			\colhead{ } &
			\colhead{ } &
			\colhead{(km s$^{-1}$)} &
			\colhead{(km s$^{-1}$)} &
			\colhead{} &
			\colhead{(min)} &
			\colhead{(MHz)} &
			\colhead{($c$)} &
			\colhead{}
		}
		\startdata
		2011/06/07 & 11226 & M2.0 & 1950 & -     & - & $\sim$20 & 12$-$30 & - & \citet{doro15} \\
		2011/09/22 & 11302 & X1.4 & 1905 & 480 & 2.4 & $\sim$2 & 10$-$90 & 0.1$-$0.4 & \citet{car13} \\
		2013/10/25 & 11882 & X2.1 & 1081 & -     & $\sim$5 & $\sim$2 & 150$-$260 & - & \citet{mor22} \\
		2014/08/25 & 12146 & M2.0 & 555  & - & -    & $\sim$2  & 62$-$82 & - & \citet{mag20} \\
		2017/09/10 & 12673 & X8.2 & 3038 & 850 & 2.9 & $\sim$5 & 30$-$60 & 0.2$-$0.25 & \citet{mor19} \\
		2019/03/20 & 12736 & C4.8 &   500 & 950 & - & $\sim$4 & 30$-$50 & 0.19 & \citet{car21} \\
		2022/03/28 & 12975 & M4.0 &  702 & 600 & $\sim$8 & $>$10 & 25$-$250 & 0.18 & \citet{mor24} \\
		2023/12/31 & 13536 & X5.0 & 2852 & 600 & - & $\sim$4 & 20$-$70 & 0.04$-$0.41 & this work 
		\enddata
\end{deluxetable*}

\begin{acknowledgments}
The authors appreciate the reviewer for his/her valuable suggestions and comments to improve the quality of this article.
We also thank Dr. Jun Dai, Profs. Yao Chen and Yihua Yan for discussions.
SDO is a mission of NASA\rq{}s Living With a Star Program. AIA data are courtesy of the NASA/SDO science teams.
STEREO/SECCHI data are provided by a consortium of US, UK, Germany, Belgium, and France.
The e-Callisto data are courtesy of the Institute for Data Science FHNW Brugg/Windisch, Switzerland.
This CME catalog is generated and maintained at the CDAW Data Center by NASA and 
The Catholic University of America in cooperation with the Naval Research Laboratory. 
SOHO is a project of international cooperation between ESA and NASA.
This work is supported by the National Key R\&D Program of China 
2021YFA1600500 (2021YFA1600502), 2022YFF0503002 (2022YFF0503000),
the Strategic Priority Research Program of the Chinese Academy of Sciences, grant No. XDB0560000,
NSFC under the grant numbers 12373065, 12333010, 12325303, 12403068, Natural Science Foundation of Jiangsu Province (BK20231510 and BK20241707), 
and the Yunnan Key Laboratory of Solar Physics and Space Science, under the number 202205AG070009.
\end{acknowledgments}

\bibliography{herb}
\bibliographystyle{aasjournalv7}

\end{document}